\documentclass[aps,preprint,groupedaddress]{revtex4}
\usepackage{graphicx}

\begin{document}

\title{Probing electron transport across a LSMO/Nb:STO heterointerface at the nanoscale}
\author{K. G. Rana, S. Parui, T. Banerjee}
 \email{T.Banerjee@rug.nl}
\affiliation{Physics of Nanodevices, Zernike Institute for Advanced Materials, University of Groningen, Nijenborgh 4, 9747 AG Groningen, The Netherlands}

\date{\today}

\begin{abstract} 
We investigate electron transport across a complex oxide heterointerface of La$_{0.67}$Sr$_{0.33}$MnO$_3$ (LSMO) on Nb:SrTiO$_3$ (Nb:STO) at different temperatures. For this, we employ the conventional current-voltage method as well as the technique of Ballistic Electron Emission Microscopy (BEEM), which can probe lateral inhomogeneities in transport at the nanometer scale. From current-voltage measurements, we find that the Schottky Barrier height (SBH) at the LSMO/Nb:STO interface decreases at low temperatures accompanied by a larger than unity ideality factor. This is ascribed to the tunneling dominated transport caused by the narrowing of the depletion width at the interface. However, BEEM studies of such unbiased interfaces, do not exhibit SBH lowering at low temperatures, implying that this is triggered by the modification of the interface due to an applied bias and is not an intrinsic property of the interface. Interestingly, the SBH at the nanoscale, as extracted from BEEM studies, at different locations in the device is found to be spatially homogeneous and similar both at room temperature and at low temperatures. Our results highlight the application of BEEM in characterizing electron transport and their homogeneity at such unbiased complex oxide interfaces and yields new insights into the origin of the temperature dependence of the SBH at biased interfaces. 
\end{abstract}

\pacs{Valid PACS appear here}
\keywords{Nb doped STO, transport}
												
\maketitle
\section{\label{sec:level1}Introduction}

Interfaces between complex oxides often exhibit unconventional transport properties normally unattainable in their individual constituents.\cite{Ohtomo}  Epitaxial Schottky interfaces involving oxide semiconductors have been investigated for their electrical transport both as a function of temperature and doping concentration of the semiconductor and are relevant for device applications.\cite{Hikita,Ruotolo,Yang,Fujii} Reliable methods employed to study transport properties are current-voltage (I-V) and capacitance-voltage (C-V) measurements and more recently the Internal Photoemission (IPE) \cite{Hikita} technique. These studies have yielded new insights, that necessitate consideration of transport mechanisms beyond that commonly employed to describe electron transport in conventional (non-oxide based) Schottky interfaces.\cite{Sze, Rhoderick} The influence of interface states and interface dipoles \cite{Minohara, HikitaPRB}, temperature and electric field dependence of the dielectric permittivity in complex oxides have to be incorporated for a complete description of the transport characteristics across such Schottky interfaces.\cite{Susaki} However, studies related to the homogeneity of the transport properties at such complex oxide interfaces, where competing electronic phases might coexist, are absent, primarily due to the limitations of the techniques used thus far. Here, using the technique of Ballistic Electron Emission Microscopy, we investigate transport properties at different regions in a La$_{0.67}$Sr$_{0.33}$MnO$_3$ (LSMO)/Nb doped SrTiO$_3$ (Nb:STO) heterostructure at the nanoscale and at different temperatures. We compare this with the current-voltage measurements performed on the same device at identical temperatures and find this to be significantly influenced by thermally activated tunneling across the interface. This is associated with a narrowing of the depletion width, due to the applied electric field, which, otherwise broadens with decreasing temperature. This enhances the probability of tunneling dominated transport in current-voltage measurements, leading to an apparent decrease in the Schottky barrier height at low temperatures and a greater than unity ideality factor in such Schottky diodes.  In BEEM, the measurements are carried out at zero bias i.e unbiased junctions and governed by the transport of hot electrons from the LSMO across the interface into Nb:STO. Here, no significant reduction in the SBH are found when the temperature is decreased.  Further, using the local probing capabilities of the BEEM, the SBH is found to be spatially homogeneous at different locations in the device both at room temperature (RT) and with decreasing temperatures. Thus, the two independent methods while probing the same interface yields new insights into the transport characteristics of the interface at the nanoscale.

\section{\label{sec:level1}Experimental Techniques}

We have used both standard \textit{I}-\textit{V} as well as the BEEM technique to study electrical transport in LSMO/Nb:STO Schottky interfaces in this work. In \textit{I}-\textit{V} measurements the bias is applied at the Metal-Semiconductor (M-S) Schottky interface (shown in Fig. 1(a)) and is varied to record the diode characteristics. We have performed these measurements both at RT and at lower temperatures upto 120 K. Theoretical models that best describe the \textit{I}-\textit{V} characteristics of these Schottky junctions are based on thermionic emission which is given by  
\begin{equation}
I=A^{*}AT^2exp(-\frac{q\phi_B}{kT})\left[exp(\frac{qV}{nkT})-1\right]
\end{equation}
where, \textit{q} is the charge of the electron, \textit{k} is the Boltzman constant, \textit{A} is the area of the diode, \textit{T} is the temperature, \textit{A*} the Richardson constant here assumed to be 156 Acm$^{-2}$K$^{-2}$ \cite{Shimizu}, $\phi_B$ is the barrier height at zero bias and \textit{n} is the ideality factor (is unity for purely thermionic emission dominated transport).

BEEM, utilizes hot electrons (few eV above $E_F$) to probe electron transport across thin metal layers deposited on a semiconducting substrate.\cite{Kaiser, Bell} The energy of the injected electrons is varied by the voltage ($V_T$) applied between the  scanning tunneling microscope (STM) tip and the metal base. As shown in Fig. 1 (b) the electrons are injected over the vacuum tunnel barrier into the metal base where they undergo inelastic (electron-electron) and elastic (structural defects, grain boundaries) scattering which reduces the number of electrons that have sufficient energy and momentum needed to overcome the Schottky barrier at the M-S interface. In BEEM, no bias is applied at the M-S interface (unbiased junctions) and electrons thus enter the conduction band of the semiconductor with their own kinetic energy. This is in contrast to the studies performed using standard $I$-$V$ where an external bias is applied at the M-S interface. The BEEM current, $I_B$, is recorded at a local area of a few nanometers, at a constant tunnel current, $I_T$, while varying $V_T$. $I_B$ depends on the thickness of the metal layer and decreases exponentially with increasing thickness.\cite{ParuiPRB, Parui,K.G.RanaSciRep} Several such spectra are collected from one location which are then averaged to obtain a representative BEEM spectrum. By placing the STM tip at different locations of the device, similar such BEEM spectra are recorded at different regions from which local Schottky barrier heights can be extracted. The onset of the BEEM current with $V_T$ gives us the SBH from the Bell-Kaiser (B-K) model.\cite{Bell} According to this model, the BEEM transmission at the interface is given by 
\begin{equation}
\frac{I_{B}}{I_T} {\hspace{0.6 mm}} {\propto} {\hspace{0.6 mm}} (V_T-\phi_B)^{2}
\end{equation}

\section{\label{sec:level1}Experimental Details}

We have used commercially available single crystalline n-type semiconducting substrate  of 0.01 wt $\%$ Nb doped SrTiO$_3$(001). Electronic characterization of such substrates were reported earlier($N$$\sim$ 8$\times$10$^{18}$ cm$^{-3}$ at RT).\cite{K.G.Rana} We have used Pulsed Laser Deposition (PLD) system to deposit a thin film of LSMO on an atomically flat TiO$_2$ terminated Nb:STO substrate.\cite{KosterB} The deposition was carried out at 750$^\circ$C in 0.35 mbar background oxygen presssure and at a laser fluence of 1.2 Jcm$^{-2}$. The LSMO film which was monitored during deposition using Reflective high energy electron diffraction (RHEED) was 13 unit cell (u.c.) thick. After deposition, the LSMO film was cooled down to RT at 100 mbar of oxygen pressure. Device structures of 250 $\mu$m$\times$1150 $\mu$m were patterned using standard UV lithography and wet etching. We have used aqua regia ( 3 parts hydrochloric acid + 1 part nitric acid) for 20 sec to etch the LSMO films to obtain these device structures. For the bottom and top contacts we used Ti(100 nm)/Au(100 nm) and Au (100 nm) respectively. For the BEEM studies, we used a modified commercial STM system from RHK technology.

\section{\label{sec:level1}Results}

The current-voltage (\textit{I-V}) characteristics of the LSMO/Nb:STO Schottky interface measured from 120 K to 300 K are shown in Fig. 2 (a). We observe i) a clear rectifying behavior at all temperatures, ii) a linear dependence of the forward current at all temperatures and iii) a shift in the onset of the current with applied bias at lower temperatures. We find that the forward characteristics tend to be less linear beyond a certain bias and is associated with the dominance of forward series resistance, which has been calculated to be around 50 $\Omega$ at RT, in our device. The reverse bias characteristics of the diode shows an increasing non-saturating current close to -1 V, which also shifts to a lower bias with decreasing temperature. These observations indicate a higher probability of electrons to flow through the interface at low temperatures and suggests the role of transport mechanisms other than the commonly used thermionic emission model. \cite{Sze,Rhoderick} The Schottky barrier height (SBH) at zero bias and the ideality factor \textit{n} can be obtained from the log \textit{I}-\textit{V} plots, by fitting the forward bias characteristics using the thermionic emission model \cite{Sze,Rhoderick} given by equation 1. The linear part of the forward characteristics of the diode obtained from  the \textit{I}-\textit{V} plot is used to obtain $\phi_B$ and \textit{n}. This is shown in Fig. 2(b). The SBH of 0.92 eV gradually decreases upto a temperature of 180 K, below which the decrease is rather abrupt (at 120 K the SBH is 0.67 eV).  The ideality factor concomitantly increases with decreasing temperature and is found to be 1.26 at the lowest temperature measured. Similar observations of a decrease in SBH and an increase in the ideality factor at low temperatures, have also been reported for different combinations of manganite/Nb:STO interfaces. \cite{Postma, Xie, Wang} 

Hot electron transmission, recorded using BEEM, across a LSMO (13 u.c.)/Nb:STO Schottky interface is shown in Fig. 3 (a) at 300 K (red) and 120 K (green). Beyond a certain tip bias, the BEEM transmission is found to increase with increasing $V_T$, while $I_T$ is held constant. The transmission shown is an average of more than 50 individual spectra recorded at a particular location in the device. By moving the STM tip across the device, similar such spectra were also recorded at different locations. As discussed earlier, the hot electrons injected across the vacuum tunnel barrier, propagate through the thin LSMO film and are collected in the conduction band of Nb:STO, provided they have the necessary energy and momentum to overcome the SBH at this interface, which is left unbiased. Electron-electron scattering processes at these energies can reduce $I_B$, which can additionally be influenced by other elastic scattering events during transmission in the LSMO film and at the M-S interface. The BEEM transmission is higher at 120 K suggesting that scattering events such as inelastic scattering or scattering due to magnons and/or phonons are reduced and also in accord with the temperature dependence of resistivity observed in LSMO films.\cite{Mercone, Renard} The onset in the BEEM transmission beyond a certain threshold voltage corresponds to the local SBH, as seen in Fig. 3(a). Using the B-K model, we obtain the SBH of the LSMO/Nb:STO interface by extrapolating the straight line so as to intersect the voltage axis in the plot of the square root of the BEEM transmission versus applied voltage as shown in Fig. 3 (b). The SBH at RT is found to be 0.87 $\pm$ 0.02 eV (error bar represents the error in the B-K fitting) and 0.83 $\pm$ 0.02 eV at 120 K. The SBH has been similarly extracted at different locations and is represented in the histogram shown in Fig. 3 (b) and (c). 

\section{\label{sec:level1}Discussions and Summary}

To understand the origin of the temperature dependence of the SBH and ideality factor from current-voltage measurements and the temperature-independence of SBH as obtained in BEEM studies, we first need to recall that the two independent techniques differ in the way they probe transport at the interface. While in the \textit{I}-\textit{V} measurements, the interface is biased that in the BEEM is not. This needs to be taken into account while choosing the appropriate transport model to explain the temperature dependence of the \textit{I}-\textit{V} data. For Schottky interfaces with complex oxides, there is large probability that electrons with energies smaller than the SBH can tunnel through the interface in what is commonly known as thermionic field emission. This occurs due to a temperature dependence of the depletion width in the oxide semiconductor. Assuming a triangular barrier \cite{Rhoderick}, the tunneling probability of the electrons within the Wentzel, Kramers and Brillouin (WKB) approximation can be written as

\begin{equation}
P = \exp\left\{\frac{-\frac{2}{3}(V_{bi})^{\frac{1}{2}}(\Delta E)^{\frac{3}{2}}}{E_{00}}\right\}
\end{equation}

where $V_{bi}$ refers to the built-in potential, $\Delta E$ indicates the amount by which the SBH is reduced for the tunneling process and $E_{00}$ is a tunneling parameter, also called the characteristic energy at 0 K given by: 

\begin{equation}
E_{00} = \frac{\hbar}{2}[\frac{N_d}{m^*\varepsilon_s}]^{1/2}
\end{equation}

where, $m^*$ is the effective mass of the electrons, $\epsilon_s$ is the permittivity of the semiconductor and $N$ is the donor concentration. $E_{00}$ is the built-in potential at a Schottky barrier which represents the transmission probability of an electron whose energy coincides with the bottom of the conduction band in Nb:STO and is equal to ${1/e}$. \cite{Rhoderick}. The current-voltage characteristics in the forward regime of the Schottky diode can thus be written as 
\begin{equation}
I = I_sexp[\frac{V}{E_0}]
\label{4}
\end{equation}where, \begin{equation}
E_{0} = nkT
\label{5}
\end{equation}
and for thermally assisted tunneling \begin{equation}
E_0 = E_{00}coth[\frac{qE_{00}}{kT}]
\label{5}
\end{equation}  

For direct tunneling $E_{00}$ $>>$ $kT$, for thermionic field emission (TFE) dominated transport $E_{00}$ $\sim$ $kT$ and for thermionic emission (TE) $E_{00}$ $<<$ $kT$. To ascertain the different contributions to transport we have calculated $E_0$ at different temperatures, using eq. 7 and show this in Fig. 4 (a). $E_{00}$ (3 meV) $<<$ $\sim$ $kT$ (25 meV) at RT, thus transport is dominated by pure thermionic emission whereas below 180 K, $E_{00}$ (3 meV) is closer to $kT$ (10 meV) and transport is thus by thermionic field emission and tunneling.\\ 
The dielectric permittivity in SrTiO$_3$ varies both as a function of temperature and electric field, thus also the depletion width in Nb:STO. Hence, the depletion width, W, can be written as  \cite{Yamamoto}
\begin{equation}
W = \frac{b\varepsilon_{0}}{qN} \cosh^{-1}\left[1+\frac{qN}{\sqrt{a}b\varepsilon_{0}}V_{bi}\right]
\end{equation}
where $N$, $V_{bi}$ are the carrier concentration in Nb:STO and the built-in potential respectively, $a$ and $b$ are temperature dependent constants in the expression that describes the temperature dependence of the dielectric permittivity, $\epsilon_{r}$, at zero electric field according to the Barrett's formula, as
\begin{equation}
\varepsilon_{r}(T) = \frac{1635}{\coth\left(\frac{44.1}{T}\right)-0.937}
\end{equation} 
$N$ is treated to be constant with decreasing temperature from Ref.\cite{K.G.Rana}.
We have calculated the depletion width using the above equations at zero applied bias and show this in Fig. 4(b). We see that the depletion width increases with decreasing temperature as expected (red curve). This cannot explain the tunneling dominated transport observed in our experiments as shown in Fig. 4(a). To understand this further, we look into the \textit{I}-\textit{V} characteristics of the LSMO/Nb:STO Schottky diode in Fig. 2(a) and see that at a temperature of 120 K, there is a shift in the onset at which the forward current in the diode sets in. This corresponds to an applied electric field, which, at such temperatures, can narrow the depletion width and enhance the prospects of thermally assisted tunneling. To confirm this, we calculate the depletion width, using eq. 8, at a bias of 0.5 V and this is shown in the blue curve in Fig. 4(b). This electric field induced reduction in the depletion width causes the electrons to tunnel through the interface at energies lower than the actual SBH and leads to an apparent decrease in the SBH and increase in the ideality factor with temperature as shown in Fig. 2(b). This additional current transport process at lower temperatures, also enhances and leads to a non-saturating current in the voltage dependence of the reverse bias current in the diode (Fig. 2(a)). \\
Electron transport and the extracted transport parameters, as the SBH, from BEEM measurements are relatively unaffected at the unbiased interface (the bias applied here is between the STM tip and the top surface of the LSMO film only), in contrast to the electric field dominated transport at the narrowed depletion width at the Schottky interface in \textit{I}-\textit{V} measurements. The local SBH as extracted from BEEM studies is reasonably similar at different locations of the device both at RT (FWHM = 0.2 eV) and at 120 K (FWHM = 0.1 eV) as has been shown in the inset of Fig. 3(b) and (c).\\
Thus, by combining two independent probes viz. current-voltage measurement and the technique of BEEM, we gain new insights into the transport characteristic in complex oxide heterostructures. We confirm that the decrease in SBH obtained from \textit{I}-\textit{V} measurements, in the temperature range studied, is primarily associated with the reduction in the depletion width (which otherwise broadens with decreasing temperatures in these oxides) with applied bias that facilitates other transport processes as tunneling. In BEEM, where transport is by hot electrons no such decrease in the local SBH with temperature is observed at the unbiased Schottky interface. Such studies of the local Schottky interface of complex oxides at the nanoscale, not demonstrated earlier, confirms not only the uniformity of the phase of the LSMO film on the Nb:STO surface but the utility of this technique for studying complex oxides where competing phases are believed to exist. \\
{\bf ACKNOWLEDGEMENTS}\\
We thank B. Noheda and T. T. M. Palstra for use of the Pulsed Laser Deposition system. Technical support from J. Baas and J. G. Holstein is thankfully acknowledged. We also acknowledge useful discussions with Y. Hikita and H. Y. Hwang. This work is supported by the Netherlands Organization for Scientific Research NWO-VIDI program and the Rosalind Franklin Fellowship\\

\clearpage

\begin{figure}[htb]
\includegraphics[scale=0.75]{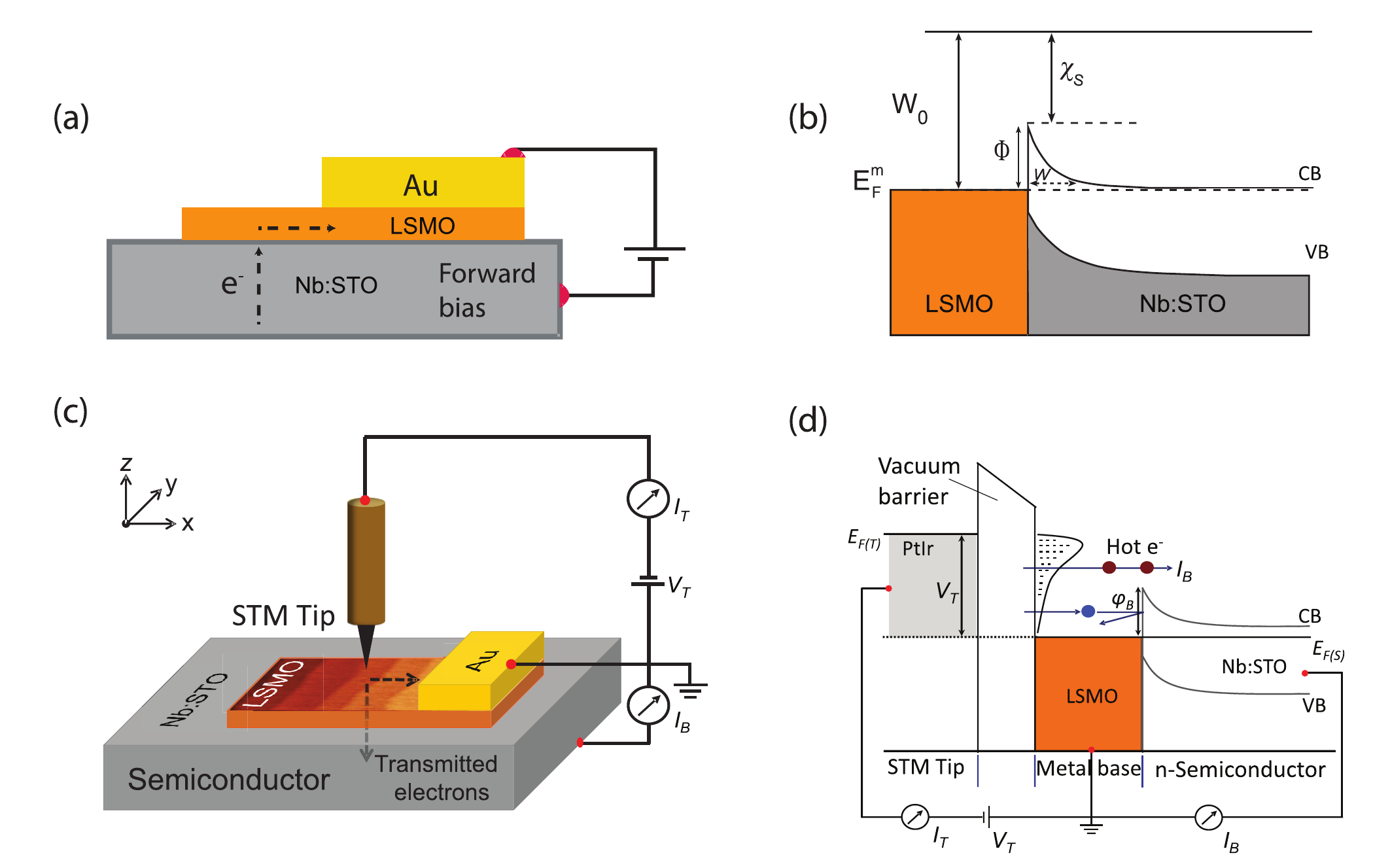}
\caption{\label{} (a) Sketch of the \textit{I}-\textit{V} measurement on a LSMO/Nb:STO device. (b) The different energy levels in the Metal and Semiconductor are shown along with the Schottky barrier height, which is defined as the difference between the work function of LSMO and electron affinity in  Nb:STO. (c) Schematic layout of the BEEM technique (not to scale). The tunnel voltage, $V_T$, is applied between the PtIr STM tip and the LSMO film, with the tunnel current, $I_T$, kept constant by feedback. The LSMO/Nb:STO interface is thus unbiased. (d) Energy band diagram of the BEEM technique shows the injected hot electron distribution. The electrons transmitted in the LSMO film, after scattering are collected in the conduction band of the Nb:STO semiconductor.}
\end{figure}

\begin{figure}[htb]
\includegraphics[scale=0.75]{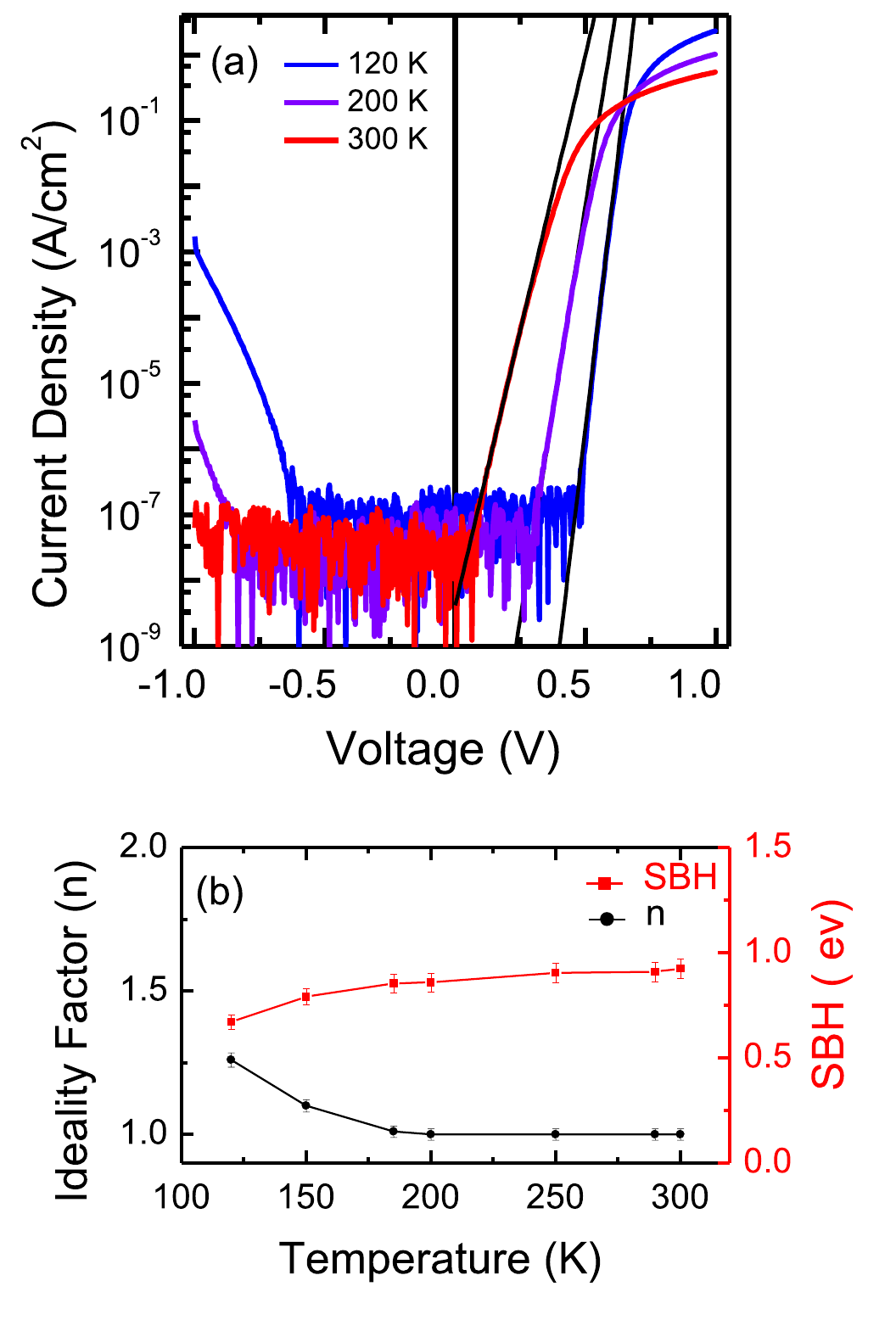}
\caption{\label{} (a) Current-Voltage \textit{I}-\textit{V} characteristics for the LSMO/Nb:STO Schottky diode shown here at Room temperature, at 200 K and 120 K. A clear rectification is observed in all cases. (b) The Schottky barrier heights and the ideality factors are extracted from (a) using eq. 1 at different measurement temperatures.}
\end{figure}

\begin{figure}[htb]
\includegraphics[scale=0.75]{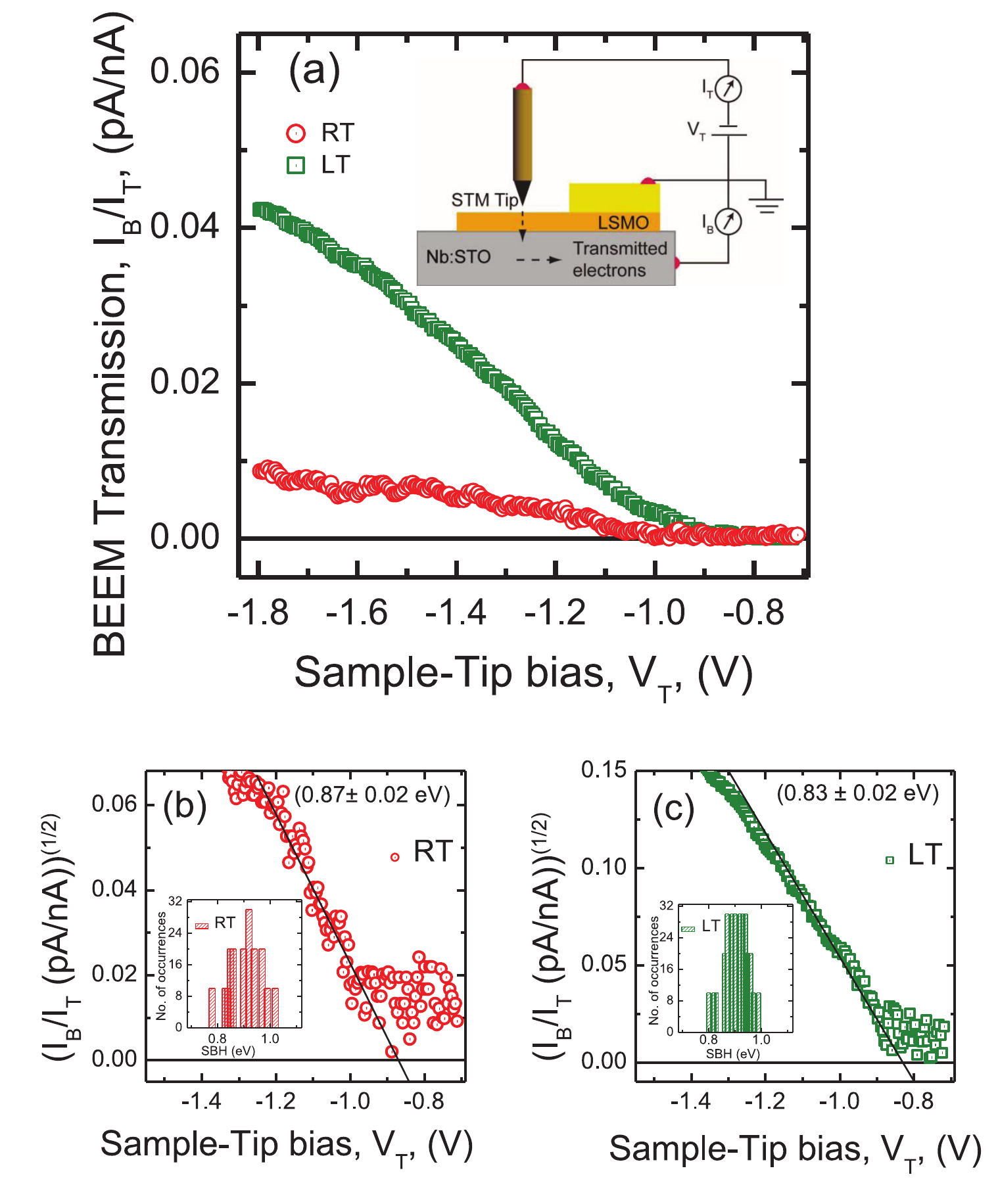}
\caption{\label{} (a) BEEM spectra for the LSMO (13 u.c.)/Nb:STO Schottky interface at RT (red) and at 120 K (green). Each spectrum is a representative of several spectra taken at the same location. (b) The extracted SBH at the LSMO (13 u.c.)/Nb:STO Schottky interface, obtained by plotting the square root of the normalized $I_B$ with $V_T$ and fitting it to the Bell-Kaiser model at RT (c) the same at 120 K, (insets) shows the distribution of Schottky barrier heights obtained at different locations in the device both at RT (b) and at 120 K (c).}
\end{figure} 

\begin{figure}[htb]
\includegraphics[scale=0.75]{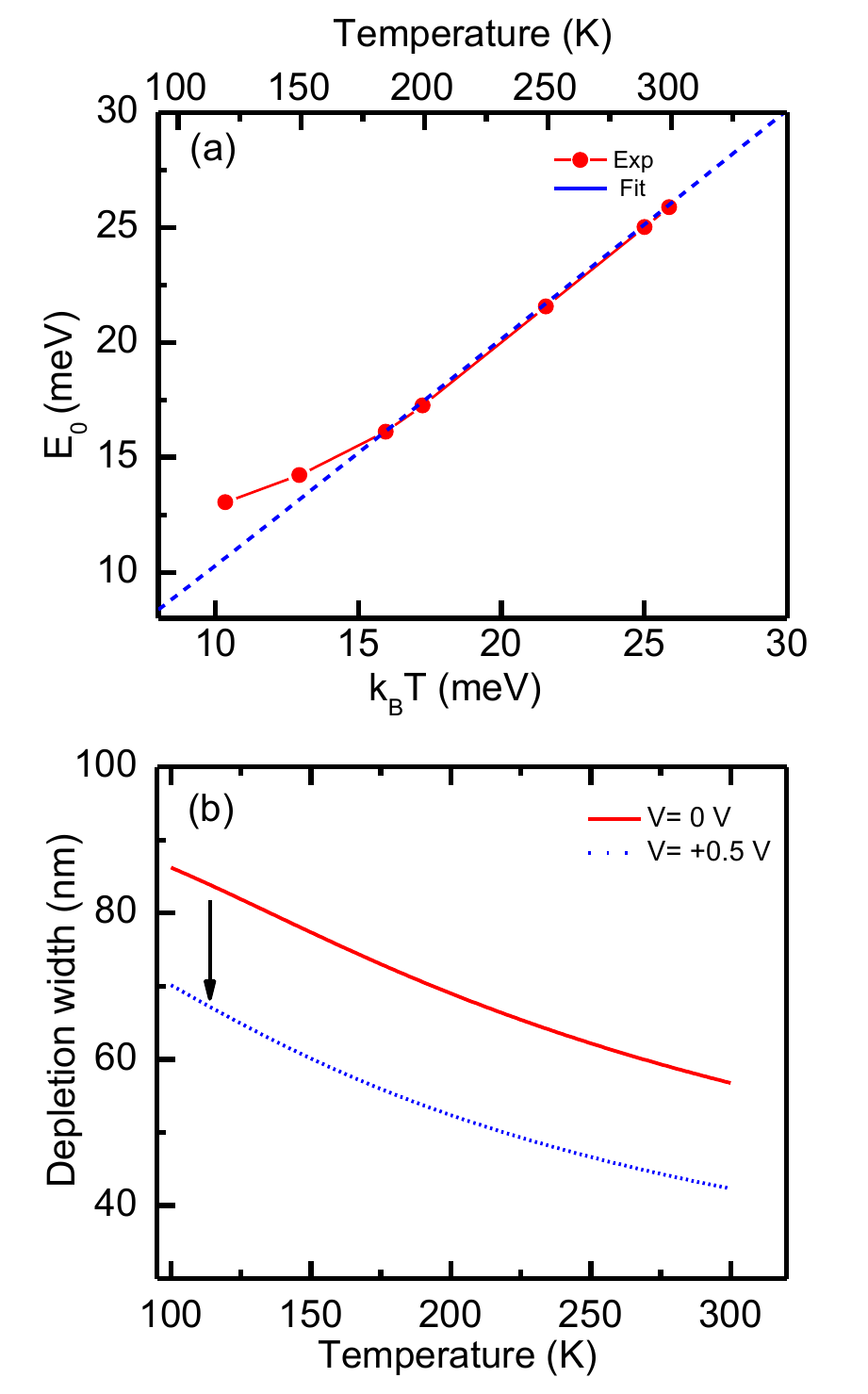}
\caption{\label{} (a) Variation of \textit{E$_0$} (tunneling parameter) with temperature. The blue dotted line represents the fit given by Eq. 7. (b) Depletion width in Nb:STO, calculated from Eq. 8 with varying temperature. The red plot is at zero bias i.e V = 0 V  whereas the blue plot is at an applied bias of 0.5 V. The blue plot shows that the depletion width in Nb:STO  reduces with the external applied field at all temperatures.}
\end{figure}

\end{document}